\documentclass{article}

\PassOptionsToPackage{square,numbers}{natbib}

\usepackage[final]{neurips_2021}

\usepackage[utf8]{inputenc} 
\usepackage[T1]{fontenc}    
\usepackage{hyperref}       
\usepackage{url}            
\usepackage{booktabs}       
\usepackage{amsfonts}       
\usepackage{nicefrac}       
\usepackage{microtype}      
\usepackage{xcolor}         
\usepackage{graphicx}
\usepackage[center]{caption}

\title{Improving the Segmentation of Pediatric Low-Grade Gliomas through Multitask Learning}

\usepackage{authblk}
\author[1,4]{Partoo Vafaeikia}
\author[2]{Matthias W. Wagner}
\author[3,5]{Uri Tabori}
\author[1,2,4,6]{Birgit B. Ertl-Wagner}
\author[1,2,4,6,7,8]{Farzad Khalvati}
\affil[1]{Neurosciences and Mental Health, The Hospital for Sick Children, Toronto, Canada}
\affil[2]{Department of Diagnostic Imaging, The Hospital for Sick Children}
\affil[3]{Division of Haematology/Oncology, The Hospital for Sick Children}
\affil[4]{Institute of Medical Science, University of Toronto\\}
\affil[5]{Department of Medical Biophysics, University of Toronto}
\affil[6]{Department of Medical Imaging, University of Toronto}
\affil[7]{Department of Mechanical and Industrial Engineering, University of Toronto}
\affil[8]{Vector Institute, Toronto, Canada}
\affil[ ]{\textit {partoo.vafaeikia@mail.utoronto.ca, matthias.wagner@sickkids.ca, uri.tabori@sickkids.ca}}
\affil[ ]{\textit {birgitbetina.ertl-wagner@sickkids.ca, farzad.khalvati@utoronto.ca}}

\begin{document}
\bibliographystyle{IEEEtranN}
\maketitle

\begin{abstract}
  Brain tumor segmentation is a critical task for tumor volumetric analyses and AI algorithms. However, it is a time-consuming process and requires neuroradiology expertise. While there has been extensive research focused on optimizing brain tumor segmentation in the adult population, studies on AI guided pediatric tumor segmentation are scarce. Furthermore, MRI signal characteristics of pediatric and adult brain tumors differ, necessitating the development of segmentation algorithms specifically designed for pediatric brain tumors. We developed a segmentation model trained on magnetic resonance imaging (MRI) of pediatric patients with low-grade gliomas (pLGGs) from The Hospital for Sick Children (Toronto, Ontario, Canada). The proposed model utilizes deep Multitask Learning (dMTL) by adding tumor's genetic alteration classifier as an auxiliary task to the main network, ultimately improving the accuracy of the segmentation results.\\
\end{abstract}

\section{Introduction}

 The development and implementation of Artificial Intelligence (AI) algorithms for brain tumors requires manual segmentation and annotation of tumors. AI and radiomics-based models developed for brain tumor classification and survival analysis often require neuroradiologists to manually segment tumors of hundreds of patients \citep{2018}. The performance of AI models can negatively be impacted by inaccurate and inconsistent manual labelling of tumor regions in medical images. Manual tumor annotation may be inaccurate due to a variety of reasons including the quality of the MRI, the neuroradiologist's experience, the intrinsic intricacy of the tumor anatomy and the inter- and intra-reader variability \citep{meier_clinical_2016}. The brain tumor segmentation (BraTS) \citep{menze_multimodal_2015} challenge is a well-known competition with the aim to evaluate state-of-the-art methods for automated tumor segmentation by providing 3D MR images of the brain of adult patients with low- and high-grade glioma and their ground truth tumor segmentation masks. Participants of this challenge in recent years \citep{henry_brain_2020,jiang_two-stage_2020,myronenko_3d_2019} have mainly focused on developing and fine tuning ensemble models based on the state-of-the-art deep learning approaches for 3D segmentation, specifically U-Net \citep{10.1007/978-3-319-24574-4_28}, to achieve the best tumor segmentation performance. Despite promising advances in MRI-based automated tumor segmentation in the adult population, there is limited work in pediatrics due to lack of adequate data. In addition, since the MRI signal characteristics of the two populations differ, it is challenging to transfer knowledge from one domain to the other using transfer learning. In an experiment, we trained a model on the BraTS adult MRI dataset and obtained a validation dice score of 0.8. This model was then applied to our pediatric dataset, which was preprocessed similar to BraTS, and achieved a dice value of 0.37. This experiment highlights the incompatibility of features in the two populations and emphasizes the significance of developing models specific to the pediatric cohort.
 
 In this paper, we propose a Machine Learning (ML) model based on deep Multitask Learning (dMTL) that automatically segments pediatric low-grade gliomas (pLGGs) in MRI scans. In dMTL, an auxiliary task (e.g., tumor classification) is considered in addition to the main task (e.g., tumor segmentation) to improve the models’ performance and generalizability ~\citep{ruder_overview_2017}. In dMTL-based networks, the layers are shared between the main and auxiliary tasks by utilizing the hard \citep{Baxter1997Bayesian/Information} or soft \citep{Duong2015Low} parameter sharing techniques. This leads to better generalization of the model by introducing an inductive bias based on various features, not only from learning the main task, but also by including auxiliary tasks that are complementary to the main task \citep{Girshick2015Fast,ruder_overview_2017}. 
 
 In pLGG, it has been shown that frequent alterations in the mitogen-activated protein kinas pathway can be identified from the molecular characterization of the tumors. These alterations have been shown to differ pLGG prognosis, and their identification allows for more focused therapies \citep{lassaletta_therapeutic_2017}. In our study, since features obtained from genetic alterations and segmentation of pLGG are interconnected, as an auxiliary task, the genetic markers classifier is added in parallel to the main network to boost segmentation results. Using dMTL, information learned from related tasks increases the model's capacity to learn a usable representation of the data, reducing overfitting and improving generalization.

\section{Methodology}
\label{methodology}
In this research, we propose an MRI-based model that provides segmentation of pLGG and at the same time, classifies genetic alteration of the tumor as an auxiliary task. To achieve this goal, fluid attenuation inversion recovery (FLAIR) MRI sequences of 311 pediatric patients treated at The Hospital for Sick Children between 2000 and 2018 were included. The ground truth tumor segmentations were provided by two neuroradiologists in consensus, and the associated genetic markers were assessed through biopsy. A U-Net based network was trained, which is the core of the state-of-the-art approaches for tumor segmentation. In order to produce classification output, we added a branch of fully connected layers at the bottleneck of the U-Net which generalizes the encoder to both segmentation and classification tasks. The SickKids Hospital Research Ethics Boards approved this retrospective study. 

\subsection{Segmentation}
Our segmentation task uses an encoder-decoder based U-Net architecture \citep{10.1007/978-3-319-24574-4_28}, where the encoder extracts image features and the decoder reconstructs the segmentation mask with a similar dimension as the input image \citep{chen_encoder-decoder_2018, chollet_xception_2017,milletari_v-net_2016}. To have the same data dimensionality and coordinate system, all images are registered to the SRI24 atlas, which is an MRI-based atlas built upon normal adult human brain anatomy \citep{rohlfing_sri24_2010}. As a result, both the input images and output segmentation masks are of size $240\times240\times155$ pixels. The architecture of the encoder is similar to that proposed by Myronenko \citep{myronenko_3d_2019}. In this network, ResNet \citep{he_deep_2016} blocks are utilized consisting of convolutions with normalization, ReLU and skip connections. At each step, the encoder downsizes image dimensions and increases the feature size by 2. The initial number of filters is 32 and the kernels are $3\times3\times3$. The structure of the decoder follows a similar pattern as the encoder and at each step, it reduces the number of features and increases the dimensionality by a factor of two. A sigmoid function is utilized in the last layer to predict the final segmentation map.

\subsection{Classification}
\begin{figure}
  \centering
  \includegraphics[scale=0.30]{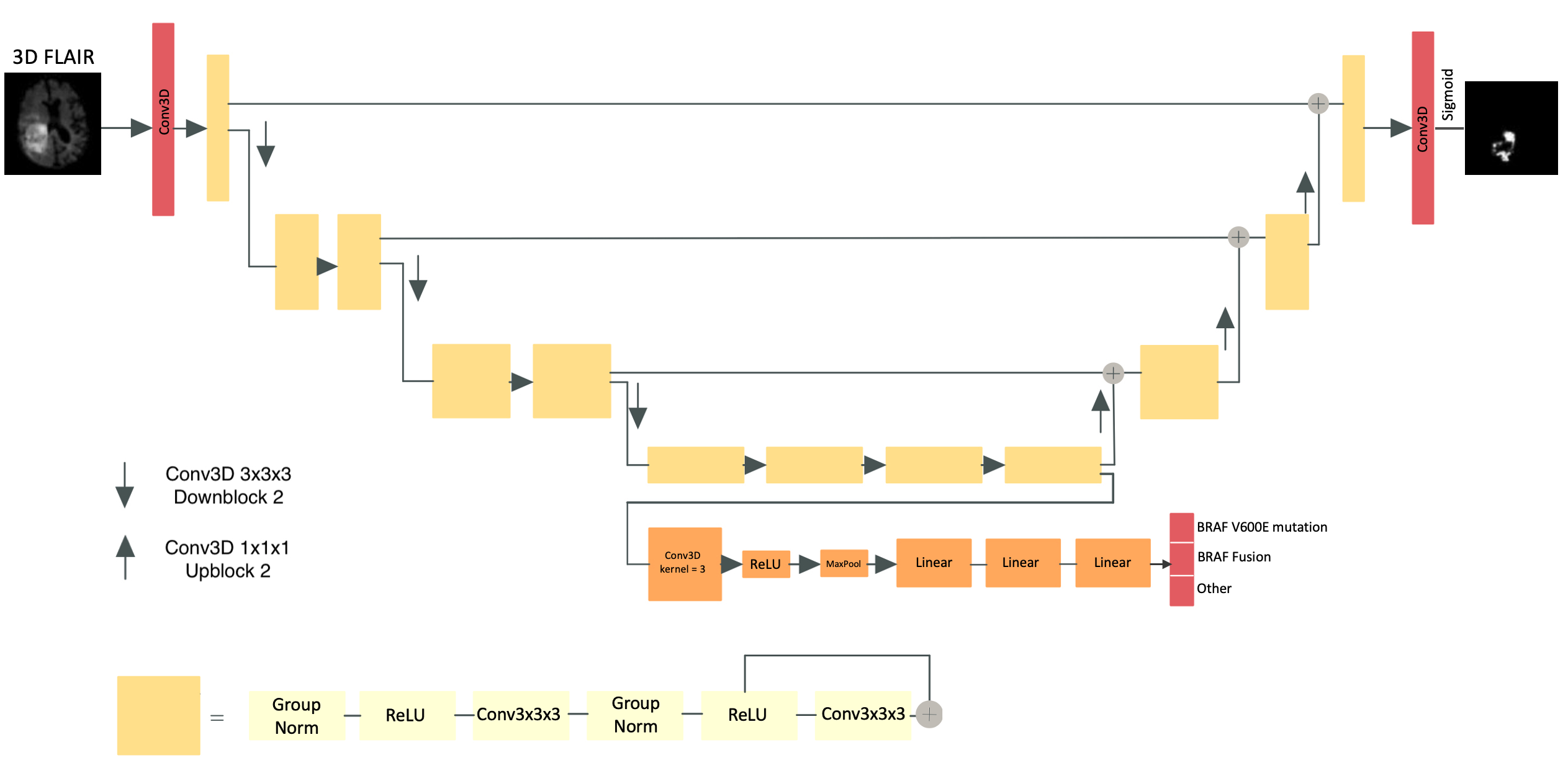}
  \caption{Model Architecture}
  \label{figure:1}
\end{figure}

PLGG is the most common brain tumor affecting children. It has recently been discovered that pLGG genetic markers have the potential to enhance diagnosis and prognosis. Currently, only invasive biopsy can provide these molecular markers and thus, there is an urgent need for non-invasive ML models to assess genetic markers from imaging data ~\cite{Wagner_2021}. Three types of genetic alterations are classified in this work: 1) BRAF V600E mutation, 2) BRAF fusion which are more commonly encountered, and 3) other molecular markers that are rare and form a third group for the purpose of this study. We added genetic marker classification as an auxiliary task to the main segmentation model. To achieve this goal, the encoder part of the segmentation U-Net was shared between the two tasks for better generalizability. An additional classification network was added to the bottleneck which consists of a branch of fully connected layers in order to output classification labels. The full model architecture is shown in Figure~\ref{figure:1}. 

\subsection{Deep Multitask Learning}
Jointly training all tasks necessitates optimization of an aggregation of individual losses into a single multi-task objective. Our loss function defined in equation~(\ref{loss}) contains two terms that measure the performance of each task; the classification task has a smaller coefficient than the segmentation task which is chosen using grid search and is proportional to the single task loss values. Thus, the model can concentrate on improving the latter (i.e., segmentation). This is a dominant approach used by previous research \citep{liao_understand_2015,sermanet_overfeat_2014,teichmann_multinet_2018}. However, since model performance is sensitive to weight selection in this scenario, we envision enhancing this definition in the next phase of our research. 
\begin{equation}
L_{dMTL} = L_{segmentation} + 0.3 * L_{classification}\label{loss}
\end{equation}
\(L_{classification}\) is a weighted cross-entropy loss and \(L_{segmentation}\) is a soft dice loss, defined in equation~(\ref{dice}) [19], which is applied to the segmentation model output, \(p_{pred}\), to match the ground truth \(p_{gt}\):
\begin{equation}
L_{segmentation} = 1 - \frac{2*\displaystyle\sum p_{pred} * p_{gt}}{\displaystyle\sum p_{pred}^2 + \displaystyle\sum p_{gt}^2 + \epsilon}\label{dice}
\end{equation}
Adam optimizer with an initial learning rate of \(lr=1e-4\) and a batch size of 4 were used in this study. Because of memory constraints, images were randomly divided into patches of  $128\times128\times112$ pixels and each experiment was run for 400 epochs. 

\section{Results}
\label{headings}
We trained two models on the same training (70\% of patients), validation (15\% of patients), and test (15\% of patients) datasets. The first model is a single task model optimized for pLGG segmentation, while the second model utilizes the proposed dMTL framework. We tried the same scenario 6 times by only changing data split in the training, validation, and test datasets. On 4 out of 6 runs, the dMTL method improved the performance of pLGG segmentation, on average, by 3\% (0.77 to 0.80) and 4\% (0.74 to 0.78) in validation and test sets, respectively, and it performed similar to single task training in the 2 remaining runs. Overall, the dMTL method improved the performance of pLGG segmentation, on average, by 2.10\% (0.767 to 0.788) and 3.0\% (0.743 to 0.773) in validation and test sets, respectively. Different level of performance improvement may be due to the discrepancy in the similarity of the training and test datasets in different data splits. The results are detailed in Table~\ref{table:1} of the appendix.

\section{Conclusion}
\label{headings}
In this research, we present a method to improve the performance of pLGG tumor segmentation task in MRI using a deep multi-task learning approach. We assessed the effect and complementarity of auxiliary representations by integrating a genetic marker classifier into the main segmentation network. Our method showed improvement in the segmentation task, on average, by 2.10\% and 3\% in validation and test datasets, respectively.

\section*{Negative Societal Impact}
Improving tumor segmentation models enables swifter development of AI algorithms for tumor characterization and prognostication. However, our model was trained on a limited number of MRI scans from one institute and was not externally validated. Larger studies with external validation are needed in the future.

\begin{ack}
This research has been supported by Canadian Cancer Society, Brain Canada Foundation, and the Canadian Institutes of Health Research (CIHR).

\end{ack}
\bibliography{sample}
\newpage
\appendix
\section{Appendix}

\begin{table}[h]
\centering
  \caption{Validation and test results for each run. Mean Dice measurements
of the proposed single and multi task model}
  \label{table:1}
  \centering
  \begin{tabular}{ccccc}
    \toprule
     \multicolumn{3}{c}{Validation Dice} & \multicolumn{2}{c}{Test Dice} \\
\cmidrule(lr){2-3} \cmidrule(lr){4-5}
    Run & Single Task & Multi Task & Single Task& Multi Task \\
    \midrule
    1  & 0.76 & \textbf{0.82} & 0.70 & \textbf{0.83}\\
    2 & 0.75  & 0.80 & 0.76 & 0.77 \\
    3 & 0.78   & 0.78 & 0.78  & 0.80  \\
    4 & 0.79 & 0.80  & 0.73  & 0.74 \\
    5 &  0.76&  0.77 & 0.80 & 0.80  \\
    6  & 0.76 & 0.76 & 0.70 &0.70   \\
    \bottomrule
    Mean Dice & 0.77 & 0.79 & 0.74 &0.77
    
  \end{tabular}
\end{table}
\end{document}